\documentclass[pre,amsmath,amssymb,twocolumn]{revtex4}

\usepackage{graphicx}
\usepackage{dcolumn}
\usepackage{bm}
\usepackage{subfigure}
\usepackage{hyperref}
\usepackage[english]{babel}

\newcommand{\mx}[1]{\mathbf{#1}}
\newcommand{\vc}[1]{\mathbf{#1}}

\begin{document}

\title{Fuzzy communities and the concept of bridgeness in complex networks}

\author{Tam\'as Nepusz}
 \email{nepusz@mit.bme.hu}
 \altaffiliation[Also at~]{KFKI Research Institute for Particle and
 Nuclear Physics of the Hungarian Academy of Sciences, Department of
 Biophysics, Budapest, Hungary and Kingston University, School of Life Sciences,
 Kingston-upon-Thames, United Kingdom.}
 \affiliation{%
Budapest University of Technology and Economics\\
Department of Measurement and Information Systems\\
H-1521 Budapest, P.O.Box 91, Hungary
}%
\author{Andrea Petr\'oczi}
\affiliation{Kingston University, School of Life Sciences\\
Kingston-upon-Thames, KT1 2EE Surrey, United Kingdom
}%

\author{L\'aszl\'o N\'egyessy}
 \affiliation{Neurobionics Research Group \\
 Hungarian Academy of Sciences - P\'eter P\'azm\'any Catholic University -
 Semmelweis University \\
 H-1094 Budapest, T\H{u}zolt\'o u. 58., Hungary
}%

\author{F\"ul\"op Bazs\'o}
 \email{bazso@sunserv.kfki.hu}
 \altaffiliation[Also at~]{Polytechnical Engineering College Subotica,
Marka Ore\v{s}kovi\'ca 16, 24000 Subotica, Serbia.}
 \affiliation{KFKI Research Institute for Particle and Nuclear Physics of the Hungarian Academy of Sciences \\
Department of Biophysics  \\
H-1525 Budapest, P.O.Box 49, Hungary
}%

\date{\today}

\begin{abstract}
We consider the problem of fuzzy community detection in networks,
which complements and expands the concept of overlapping community
structure. Our approach allows each vertex of the graph to belong to multiple
communities at the same time, determined by exact numerical membership
degrees, even in the presence of uncertainty in the data being analyzed. We created
an algorithm for determining the optimal membership degrees with
respect to a given goal function. Based on the membership degrees, we
introduce a new measure that is able to identify outlier vertices
that do not belong to any of the communities, bridge vertices
that belong significantly to more than one single community, and regular
vertices that fundamentally restrict their interactions within their
own community, while also being able to quantify the centrality of a vertex
with respect to its dominant community. The method can also be used for
prediction in case of uncertainty in the dataset analyzed. The number of
communities can be given in advance, or determined by the algorithm itself
using a fuzzified variant of the modularity function. 
The technique is able to discover the fuzzy community structure of different
real world networks including, but not limited to social networks, scientific
collaboration networks and cortical networks with high confidence.
\end{abstract}


\maketitle

\section{Introduction}

Recent studies revealed that graph models of many real world phenomena
exhibit an overlapping community structure, which is hard to grasp with the
classical graph clustering methods where every vertex of the graph belongs to
exactly one community \cite{palla05}. This is especially true for social
networks, where it is not uncommon that individuals in the network belong to more
than one community at the same time. Individuals who connect groups in the
network function as ``bridges'', hence the concept of ``bridge'' is defined as the
vertices that cross structural holes between discrete groups of people
\cite{burt92}. It is therefore important to define a quantity that measures
the commitment of a node to several communities in order to obtain a more
realistic view of these networks.

The intuitive meaning of a bridge vertex may differ in different kinds of
networks that exist beyond sociometrics. In protein interaction networks,
bridges can be proteins with multiple roles. In cortical networks containing
brain areas responsible for different modalities (for instance, visual and
tactile input processing), the bridges are presumably the areas that take part
in the integration and higher level processing of sensory signals. In word
association networks, words with multiple meanings are likely
to be bridges \footnote{Bridges described in this paper are not to be confused
with the concept of cut edges which are sometimes also referred as bridges in
classical graph theory. Articulation points (vertices whose removal disconnects the
remaining subgraph) bear more similarity to the concept of bridges described in
this paper, but not all bridge vertices are articulation points. From the
structural perspective the concept of bridge and bridgeness may be considered
as a generalization of the notion of articulation point, suitably tailored to
the problem of community detection.}.
The state-of-the-art overlapping community detection algorithms
\cite{reichardt04,palla05,capocci05,zhang07} are not able to quantify the notion of
bridgeness, while other attempts at quantifying it (e.g., the participation
index \cite{guimera05}) are only concerned with non-overlapping communities.

\begin{figure}[tb]
\subfigure[~]{
\label{fig:toy_graph}
\includegraphics[width=0.22\textwidth]{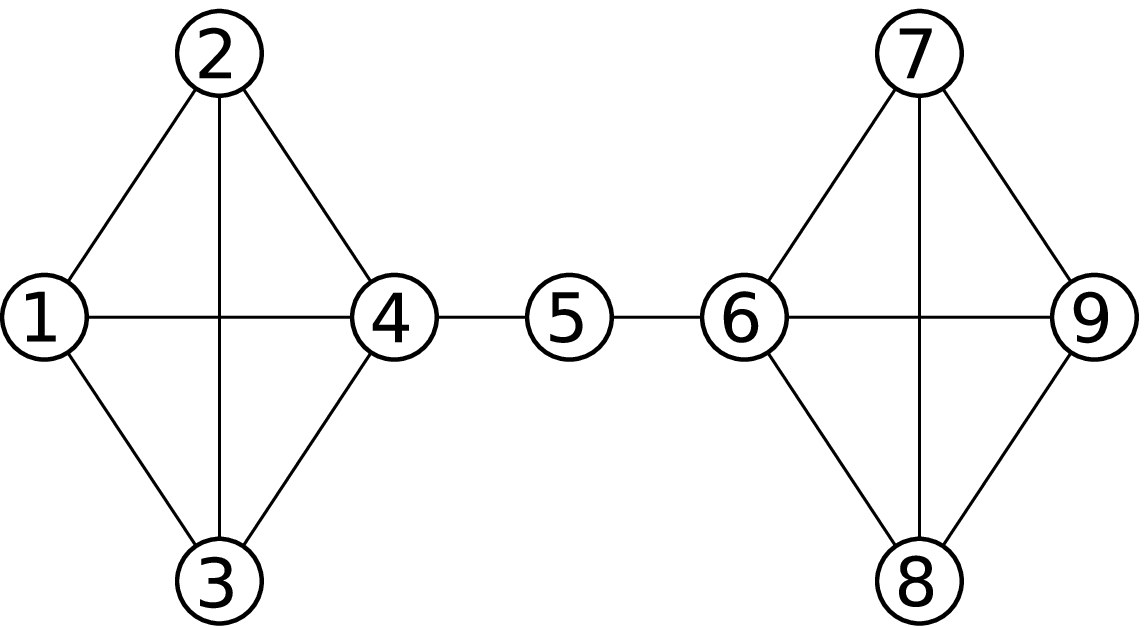}
} \hfill
\subfigure[~]{
\label{fig:toy_graph_dendro}
\includegraphics[width=0.22\textwidth]{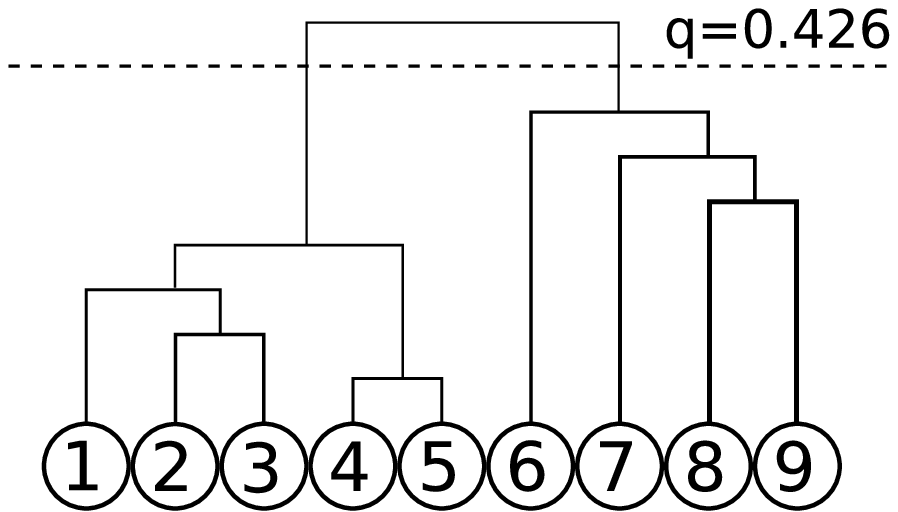}
}
\caption{Panel $(a)$: a simple graph that is unable to be partitioned into two
communities without allowing overlaps or outliers. Panel $(b)$: the dendrogram
of the graph as calculated by the greedy modularity optimization algorithm
\protect\cite{clauset04}. The dashed line denotes the level where the dendrogram
should be cut in order to reach the maximal modularity (denoted by $q$).
}
\end{figure}

To emphasize the importance of bridge vertices in community detection and to
illustrate the concept, we take a simple graph shown on
Fig.~\ref{fig:toy_graph} as an example.  A visual inspection of this graph most
likely suggests two densely connected communities,
with vertex 5 standing somewhere in between, belonging to both of them at the same
time. One may argue that vertex 5 itself forms a separate community, but a
community with only a single node is usually not meaningful (and we can also
easily add more edges connecting the two communities to vertex 5 to emphasize
its sharedness). This property of vertex 5 is not revealed by any classical
community detection algorithm without accounting for overlaps or outliers.

Hierarchical algorithms build a dendrogram from the vertices by joining them to
communities one by one (or starting from the opposite direction, splitting the
graph into two subcommunities, then splitting the subcommunities again until every
vertex forms a single community). For instance, the modularity optimization
algorithm of Clauset \emph{et al} \cite{clauset04} repeatedly merges individual
vertices or already created communities to form bigger ones in a way that
greedily maximizes the modularity of the achieved partition (for the definition
of modularity, see \cite{newman04} or Eq.~\ref{eq:modularity}). Using this
algorithm, vertex 5 was merged to vertex 4 right at the first step of the
algorithm, misleadingly suggesting that they can not be separated from each
other. The complete dendrogram is shown on Fig.~\ref{fig:toy_graph_dendro}.

A better solution can be achieved by applying the clique percolation method
(CPM) of Palla \emph{et al} \cite{palla05}, which is also able to discover
overlapping communities. In this case, vertex 5 was classified as an
outlier (a vertex that does not belong to any community). This result
stands closer to our visual inspection and clearly underlines the fact that
in many cases, we should not assume that a vertex belongs to one and only one
community in the graph. However, vertex 5 is not an outlier in the sense that
removing it from the network would result in two disconnected components. Vertex
5 is an integral part of the network, serving as the only connection between
two densely connected subgroups.


\section{Methods}

\subsection{Fuzzy community detection as a constrained optimization problem}

The objective of classical community detection in networks is to partition the
vertex set of the graph $G(V, E)$ into $c$ distinct subsets in a way that puts
densely connected groups of vertices in the same community. $c$ can either be
given in advance or determined by the community detection algorithm itself. For
the time being, let us assume that $c$ is known. In this case, a convenient
representation of a given partition is the \emph{partition matrix} $\mx{U} =
\left[u_{ik}\right]$ \cite{bezdek81}. $\mx{U}$ has $N = |V|$ columns and $c$
rows, and $u_{ik} = 1$ if and only if vertex $k$ belongs to the $i$th subset
in the partition, otherwise it is zero.  From the definition of the partition,
it clearly follows that $\sum_{i=1}^c u_{ik} = 1$ for all $1 \le k \le N$.
The size of community $i$ can then be calculated as $\sum_{k=1}^N u_{ik}$,
and for any meaningful partition, we can assume that $0 < \sum_{k=1}^N u_{ik} < N$.
These partitions are traditionally called \emph{hard} or \emph{crisp}
partitions, because a vertex can belong to one and only one of the
detected communities \cite{bezdek81}.

The generalization of the hard partition follows by allowing $u_{ik}$ to
attain any real value from the interval $[0, 1]$. The constraints imposed on
the partition matrix remain the same \cite{ruspini70}:
\begin{subequations}
\label{eq:constr}
\begin{eqnarray}
u_{ik} \in [0, 1] \mbox{~for all~} 1 \le i \le c, 1 \le k \le N \label{eq:constr0} \\
\sum_{i=1}^c u_{ik} = 1 \mbox{~for all~} 1 \le k \le N \label{eq:constr1} \\
0 < \sum_{k=1}^N u_{ik} < N \mbox{~for all~} 1 \le i \le c. \label{eq:constr2}
\end{eqnarray}
\end{subequations}
Eq.~\ref{eq:constr1} simply states that the total membership degree for each vertex
must be equal to 1. Informally, this means that vertices have a total membership
degree of 1, which will be distributed among the communities.
Eq.~\ref{eq:constr2} is the formal description of a simple requirement: we are
not interested in empty communities (to which no vertex belongs to any extent),
and we do not want all vertices to be grouped into a single community. Partitions
of this type are called \emph{fuzzy partitions}. The fuzzy membership degrees
for a given vertex can be thought about as a trait vector that describes some
(possibly nonobservable) properties of the entity which the vertex represents in a
compact manner. Trait-based graph models have already been suggested as models
for complex networks \cite{zalanyi03}.

Since the groundbreaking work of Dunn \cite{dunn73} and Bezdek \cite{bezdek81}
on the fuzzy $c$-means clustering algorithm, many methods have been developed
to search for fuzzy clusters in multi-dimensional datasets. For an overview of
these methods, see Bezdek and Pal \cite{bezdek92}. However, these methods
usually require a distance function defined in the space the data belong to,
therefore it is impossible to apply them to graph partitioning directly, except
in cases where the vertices of the graph are embedded in an $n$-dimensional
space.  A recent paper of Zhang \emph{et al} \cite{zhang07} discusses a
possible embedding of the vertices of an arbitrary graph into an
$n$-dimensional space using spectral mapping in order to utilize the fuzzy
$c$-means algorithm on graphs. They were able to identify meaningful fuzzy
communities in several well-known test graphs (e.g., the Zachary karate club
network \cite{zachary77} and the network of American college football teams
\cite{girvan02}), but the eigenvector calculations involved in the algorithm
render it computationally expensive to use on large networks. 

To overcome the need of spatial embedding, we propose a different approach based
on vertex similarities. We observe that a meaningful partition (let it be hard or
fuzzy) should group vertices that are somehow similar to each
other in the same community. It is reasonable to assume that an edge
between vertex $v_1$ and $v_2$ implies the similarity of $v_1$ and $v_2$, and
likewise, the absence of an edge implies dissimilarity. Let us assume that
we have a function $s(\mx{U}, i, j)$ that satisfies the following criteria:

\begin{enumerate}
\item $s(\mx{U}, i, j) \in [0, 1]$
\item $s(\mx{U}, i, j)$ is continuous and differentiable for all $u_{ij}$.
\item $s(\mx{U}, i, j) = 1$ if the membership values of $v_i$ and $v_j$
suggest that they are as similar as possible.
\item $s(\mx{U}, i, j) = 0$ if the membership values of $v_i$ and $v_j$
suggest that they are completely dissimilar (there is no chance that they
belong to the same community).
\end{enumerate}

Let us call such $s(\mx{U}, i, j)$ a \emph{similarity function}, and for
the sake of simplicity, we simply denote it by $s_{ij}$ from now on (not
emphasizing its dependence on $\mx{U}$).  Suppose we have a prior assumption
about the actual similarity of the vertices, denoted by $\tilde{s}_{ij}$
for $v_i$ and $v_j$. This leads us to the following equation, which measures
the fitness of a given partition $\mx{U}$ of graph $G(V, E)$ by quantifying
how precisely it approximates the prescribed similarity values with
$s_{ij}$:
\begin{equation}
D_G(\mx{U}) =
\sum_{i=1}^{N} \sum_{j=1}^{N} w_{ij} \left( \tilde{s}_{ij} - s_{ij} \right)^2,
\label{eq:goal_function}
\end{equation}
where $w_{ij}$'s are optional weights and $N = |V|$ is the number of vertices in
the graph. For the sake of notational simplicity, we also introduce the
matrices $\mx{W} = \left[w_{ij}\right]$, $\mx{S}(\mx{U}) =
\left[s_{ij}\right]$ and $\mx{\tilde{S}} =
\left[\tilde{s}_{ij}\right]$. From now on, we assume that
$\mx{\tilde{S}} = \mx{A}_G$, the adjacency matrix of the graph, in
concordance with our assumption that the similarity of connected vertex pairs
should be close to 1 and the similarity of disconnected vertex pairs should
be close to zero. The only thing left is to precisely define a similarity
function $s_{ij}$ that satisfies the conditions prescribed above. The definition
we used was the following:
\begin{equation}
s_{ij} = \sum_{k=1}^c u_{ki} u_{kj}
\label{eq:similarity_definition}
\end{equation}
It easily follows that $\mx{S}(\mx{U}) = \left[ s_{ij} \right] = \mx{U}^T \mx{U}$
\footnote{The matrix form of this problem bears some similarity with the Cholesky
decomposition. For positive weights, $D_G(\mx{U})$ is zero if and only if
$\mx{\tilde{S}}=\mx{U}^T \mx{U}$. This would be easy to solve if $\mx{U}$
was an $n \times n$ matrix (meaning that the number of communities $c$ is equal
to the number of vertices $n$), and $\mx{\tilde{S}}$ was symmetric and
positive-definite. Since none of these conditions hold, all that we can do is
to minimize the difference between $\mx{\tilde{S}}$ and $\mx{U}^T \mx{U}$ by
finding an appropriate $\mx{U}$.}.

In summary, the community detection problem in this framework boils down
to the optimization of $D_G(\mx{U})$ defined in Eq.~\ref{eq:goal_function}: we
must find $\mx{U}$ that minimizes $D_G(\mx{U})$ while satisfying the
conditions of Eq.~\ref{eq:constr}. The number of clusters $c$, the weight
matrix $\mx{W}$ and the desired similarities $\mx{\tilde{S}}$ are given in
advance (the latter one most commonly equals to the adjacency matrix
$\mx{A}_G$). This is a nonlinear constrained optimization problem. Although
there exist a set of necessary conditions that restrict the set of possible
$\mx{U}$'s worth evaluating \cite{karush39,kuhn51}, the computationally most
feasible approach to optimize $D_G(\mx{U})$ is to use a gradient-based
iterative optimization method (e.g., simulated annealing). The equality
constraints in Eq.~\ref{eq:constr1} can be incorporated into the goal function
by Lagrangian multipliers $\vc{\lambda} = \left[ \lambda_1, \lambda_2, \dots
\lambda_N \right]$, resulting in the following modified goal function:
\begin{eqnarray}
\tilde{D}_G(\mx{U}, \vc{\lambda}) & = \displaystyle
\sum_{i=1}^{N} \sum_{j=1}^{N} w_{ij} \left( \tilde{s}_{ij} - s_{ij} \right)^2 + \nonumber \\
& \displaystyle \sum_{i=1}^N \lambda_i \left( \sum_{k=1}^c u_{ki} - 1\right),
\label{eq:modified_goal_function}
\end{eqnarray}
The modified goal function compactly encodes the original goal function and the
constraints, since $\nabla_{u_{ij}} \tilde{D}_G(\mx{U}, \vc{\lambda}) = 0$ (for all
$1 \le i \le c$ and $1 \le j \le N$) ensures that we are at a stationary point of
the goal function, and $\nabla_{\vc{\lambda}} \tilde{D}_G(\mx{U}, \vc{\lambda}) = 0$
ensures that we satisfy the conditions of Eq.~\ref{eq:constr1}. Therefore,
stationary points of Eq.~\ref{eq:modified_goal_function} will also be stationary
points of Eq.~\ref{eq:goal_function} and they do not violate Eq.~\ref{eq:constr1}.

To employ a gradient-based iterative optimization method, we need the derivatives
of the goal function with respect to $u_{kl}$. First we note that
\begin{equation}
\frac{\partial s_{ij}}{\partial u_{kl}} =
\frac{\partial}{\partial u_{kl}} \left( u_{ki} u_{kj} \right)
\end{equation}
which is zero, except when $i=l$ or $j=l$:
\begin{equation}
\frac{\partial s_{ij}}{\partial u_{kl}} =
\left\{
\begin{array}{ll}
2 u_{kl} & \mbox{if~}i=l \wedge j=l \\
u_{ki} & \mbox{if~}i \ne l \wedge j=l \\
u_{kj} & \mbox{if~}i=l \wedge j \ne l \\
0 & \mbox{if~}i \ne l \wedge j \ne l
\end{array}
\right.
\end{equation}
The partial derivative of $\tilde{D}_G(\mx{U}, \vc{\lambda})$ with respect to
$u_{kl}$ is therefore
\begin{eqnarray}
\frac{\partial \tilde{D}_G}{\partial u_{kl}} 
& = & \displaystyle - 2 \sum_{i=1}^N w_{il} ( \tilde{s}_{il} - s_{il} ) u_{ki} \nonumber \\
& & \displaystyle - 2 \sum_{j=1}^N w_{lj} ( \tilde{s}_{lj} - s_{lj} ) u_{kj} +
\lambda_l
\label{eq:goal_function_derivative}
\end{eqnarray}

Let $e_{ij} = w_{ij} \left( \tilde{s}_{ij} - s_{ij} \right)$. Summing the partial
derivatives for $k=1, 2, \dots c$, making them equal to zero and substituting
Eq.~\ref{eq:constr1} back where appropriate leaves us with:
\begin{eqnarray}
\lambda_l & = & \displaystyle \frac{2}{c} \sum_{i=1}^N (e_{il} + e_{li})
\label{eq:lambda}
\end{eqnarray}

The substitution of Eq.~\ref{eq:lambda} into Eq.~\ref{eq:goal_function_derivative}
yields one component of the goal function's gradient vector:
\begin{eqnarray}
\frac{\partial \tilde{D}_G}{\partial u_{kl}} 
& = & 2 \sum_{i=1}^N (e_{il}+e_{li}) \left( \frac{1}{c} - u_{ki} \right)
\label{eq:gradient}
\end{eqnarray}

The simplest gradient-based algorithm for finding a local minimum of $\tilde{D}_G$
is then the following:

\begin{enumerate}
\item Start from an arbitrary random partition $\mx{U}^{(0)}$ and let $t = 0$.
\item Calculate the gradient vector of $\tilde{D}_G$ according to
  Eq.~\ref{eq:gradient} and the current $\mx{U}^{(t)}$.
\item If $\max_{k,l} |\frac{\partial \tilde{D}_G}{\partial u_{kl}}|<\varepsilon$,
  stop the iteration and declare $\mx{U}^{(t)}$ a solution.
\item Otherwise, calculate the next partition in the iteration with the
  following equation:
  \begin{equation}
u_{ij}^{(t+1)} = u_{ij}^{(t)} + \alpha^{(t)} \frac{\partial \tilde{D}_G}{\partial u_{ij}}
  \end{equation}
  where $\alpha^{(t)}$ is a small step size constant chosen appropriately.
\item Increase $t$ and continue from step 2.
\end{enumerate}

$\alpha^{(t)}$ can be determined by a line search towards the direction
defined by the gradient vector, it can be adjusted iteratively
according to some simulated annealing schedule (see \cite{nourani98} for
a comparison of strategies), or it can be made adaptive from iteration to
iteration by checking the difference of the values of the goal function in
the last few steps: the step size can be increased if the value of the goal
function decreased, and it must be decreased if the value of the goal
function increased. We must also make sure that the
procedure does not end up in a saddle point or a local maximum of $D_G(\mx{U})$
\footnote{Local maxima are easy to avoid by choosing an $\alpha^{(t)}$ that
always decreases the value of the goal function in the next step. Saddle points and
not too deep local minima can be avoided by randomly mutating the acquired
solution and see if the iteration converges back to the original, unmutated
solution.}.

According to our simulations, the quality of
the result is not affected by the initial membership degrees, but the
speed of convergence is. In the extreme case, if we choose all $u_{ij}$ to
be equal to $1/c$, all the gradients will be zero (see
Eq.~\ref{eq:gradient}), therefore it is suggested to use a randomized
initial partition matrix. The best results can be achieved by choosing
the initial membership degrees from a uniform distribution while still
satisfying the sum constraints. Uniformity with respect to the constraints is
not straightforward to achieve. The intuitive approach is to choose a
random number from the interval $[0, 1]$ for every $u_{ij}$ and divide
them with their respective column sums to satisfy Eq.~\ref{eq:constr1}.
However, this method is biased towards membership vectors describing
vertices equally participating in every community.  The proper way to sample
from all possible membership vectors is to draw every vector from a
Dirichlet distribution with order $c$ and $\vc{\alpha} = \left[ 1, 1, \dots, 1
\right]$ where $\vc{\alpha}$ has $c$ coordinates.  Such a distribution can be
generated by drawing $c$ independent random samples from gamma distributions
each with shape and scale parameters equal to 1, and dividing each variable
with the sum of all of them \cite{devroye86}.

\begin{figure}
\includegraphics[width=0.45\textwidth]{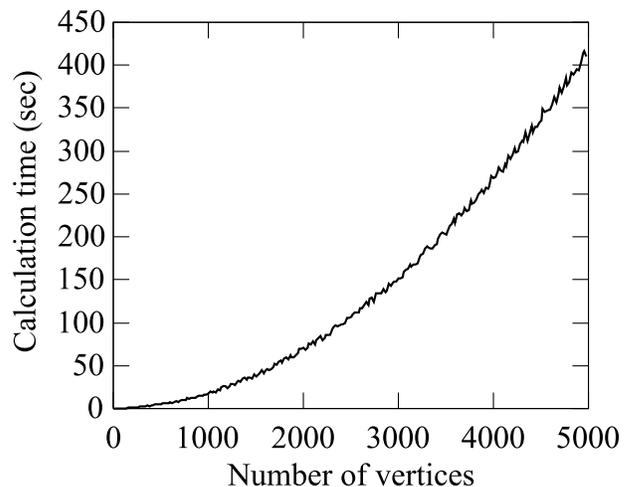}
\caption{Running time of our algorithm as a function of the number of
vertices in a graph with 4 communities. The hardware used for calculation
was an 1.83 GHz Intel Core Duo MacBook. Fitting $f(x) = ax^b$ resulted in
the parameters $a = 2.3 \times 10^{-5} \pm 1 \times 10^{-6}$ and
$b = 1.968 \pm 4.24 \times 10^{-3}$ (standard deviation from the fitted
curve = 2.583), confirming our reasoning on the quadratic running time
of the algorithm.\label{fig:time_complexity}}
\end{figure}

With $N$ vertices and $c$ communities, the time complexity of calculating the
initial membership is $O(Nc)$, calculating the gradient vectors in each step
is $O(N^2c)$, choosing the maximum gradient component for each vertex
is $O(Nc)$ and calculating the next partition matrix is $O(Nc)$, assuming
that the step size can be chosen in $O(1)$ (which is true for simulated annealing
strategies or adaptive step sizes based on the decline of the goal function
between subsequent steps). This results in an overall time complexity of
$O(N^2ch)$, where $h$ is the number of steps necessary for the algorithm to
terminate, meaning that the calculation time is expected to scale quadratically
with the number of vertices if $N \gg c$, which is confirmed by our measurements.
The time complexity of our implementation (Fig.~\ref{fig:time_complexity}) is
slightly worse than that of spectral methods, where an almost linear time complexity can be achieved by, e.g., using the implicitly restarted Arnoldi method
\cite{lehoucq96} to compute some of the largest eigenvectors.

For the sake of completeness, we show that $\mx{U}^{(t+1)}$ remains a partition
matrix if $\mx{U}^{(t)}$ was a partition matrix. We recall that a partition matrix
satisfies Eq.~\ref{eq:constr0} and Eq.~\ref{eq:constr1}. In the first step, we
choose $\mx{U}^{(0)}$ that satisfies Eq.~\ref{eq:constr2}. The persistence of
Eq.~\ref{eq:constr0} and Eq.~\ref{eq:constr2} is straightforward if we always
keep $\alpha^{(t)}$ low enough, so we only have to prove the persistence of
Eq.~\ref{eq:constr1}:

\begin{eqnarray}
\sum_{i=1}^c u_{ik}^{(t+1)} & = & \sum_{i=1}^c u_{ik}^{(t)} +
\sum_{i=1}^c \alpha^{(t)} \frac{\partial \tilde{D}_G}{\partial u_{ik}^{(t)}} \nonumber \\
& = & 1 + 2 \alpha^{(t)} \sum_{i=1}^c \sum_{j=1}^N (e_{jk}+e_{kj}) \left( \frac{1}{c} - u_{ij}^{(t)} \right) \nonumber \\
& = & 1 + 2 \alpha^{(t)} \sum_{j=1}^N (e_{jk}+e_{kj}) \left( 1 - \sum_{i=1}^c u_{ij}^{(t)} \right) \nonumber \\
& = & 1
\end{eqnarray}

\subsection{The concept of bridgeness}

\label{bridgeness}
One of the advantages of fuzzy community detection is that it enables us to
analyze to what extent a given vertex is shared among different communities.
This measure is called \emph{bridgeness}. Intuitively, a vertex that belongs to
only one of the communities has zero bridgeness, while a vertex that belongs to
all of the communities exactly to the same extent has a bridgeness of 1. We
define the bridgeness of a vertex $v_i$ as the distance of its membership
vector $\vc{u}_i = \left[ u_{1i}, u_{2i}, \dots, u_{ci} \right]$ from the
reference vector $\left[ \frac{1}{c}, \frac{1}{c}, \dots, \frac{1}{c} \right]$
in the Euclidean vector norm \footnote{Other vector norms are also conceivable
with different normalization factors to make the result span over the interval
$[0, 1]$.}, inverted and normalized to the interval $[0, 1]$ as follows:

\begin{equation}
b_i = 1 - \sqrt{\frac{c}{c-1} \sum_{j=1}^c \left(u_{ji}-\frac{1}{c}\right)^2}
\end{equation}

Note that $b_i$ attains its theoretical maximum when $v_i$ belongs to all of the
communities exactly with the same membership degree, therefore it is possible
that in this case, $v_i$ is more likely to be an outlier in the graph (a vertex
belonging to none of the communities) rather than a bridge. To distinguish
outliers and real bridges, one should also look at the centrality measures of
the node: high centrality supports the assumption that the vertex is effectively
a bridge, because despite its central role in the network, the algorithm was not
able to assign it to a single community. Low centrality may mean that the algorithm
strived to make the vertex dissimilar from almost all other vertices, therefore it
made it belong to all the communities. The simplest measure that incorporates
centrality and bridgeness score into a single number is simply defined
as the product of the degree and the bridgeness of the node, and will be called
\emph{degree-corrected bridgeness} from now on. Other centrality measures
(e.g. betweenness centrality, closeness centrality or eigenvector centrality)
can also be used. More sophisticated centrality measures take into account that
several networks contain vertices that have a crucial role but a relatively low
degree (e.g. metabolic networks, as shown in \cite{guimera05}).
We also suggest to plot a chosen centrality measure versus the bridgeness score
for each vertex to visually aid the selection of bridge vertices and outliers.
An example of this kind of plot will be shown in Section \ref{results} on
Fig.~\ref{fig:cortex_degree_br_plot}.

Bridgeness can either be used in benchmarks to assess how sensitive the
algorithm is to structural overlaps, or in the analysis of real data to gain
information about the roles of the vertices in the network. Vertices with high
centrality and bridgeness scores close to zero are likely to be in the cores of the
communities, while bridgeness scores close to one with a high centrality suggest
vertices standing in a bridgelike position between communities. In this sense,
substracting the bridgeness score from 1 and multiplying it by an appropriate
centrality measure results in a measure of the centrality of the vertex with respect
to its own communities in the network, similarly to the measure introduced in
\cite{newman06}. Benchmark results and the application of bridgeness in data
analysis is presented in Section \ref{results}.

\section{Parametrization of the algorithm}

\label{distance_based_relaxation}

At first glance, it may seem difficult to select the appropriate value for
each parameter of the algorithm described in the previous section.
However, most of these parameters have reasonable default values that can
be used in most cases. The only exception is the number of clusters $c$,
for which we will describe a simple process to identify its most suitable
value. In this section, we explain the key ideas one should consider when
choosing the appropriate values for the parameters.

\subsection{Choosing the number of communities}

The first and most important parameter of the method is $c$, defining the
number of communities the algorithm tries to discover in the network.  This
parameter is the keystone of most community detection algorithms, and
determining $c$ in a self-consistent way without human intervention is
definitely a complicated problem. Spectral methods rely on the largest
eigenvalues of the adjacency matrix $\mx{A}_G$ or the smallest eigenvalues of
the Laplacian matrix $\mx{L}_G = \mx{A}_G - \mx{D}_G$ (where $\mx{D}_G$ is a
diagonal matrix with diagonal elements $k_i$, the degrees of the vertices)
to define the number of communities, but this is usually done by visual
inspection, and since the eigenspectrum of most networks found in real
applications resemble a straight line instead of a step function, choosing $c$
is not free of subjective elements. For instance, the number of
eigenvalues of the Laplacian matrix of a graph that are close to zero are often
used as the value of $c$, but this only replaces the value of $c$ with another
parameter: a threshold level that decides which eigenvalues are considered to
be close to zero. The threshold is then chosen manually.

In order to get rid of the human intervention needed to choose $c$ based on the
eigenvalues, we propose a different, divisive approach which also spares some
computation in the early stage of the algorithm.  Initially, we compute a fuzzy
bisection of the graph by setting $c=2$. After that, whenever the optimization
gets stuck in a local minimum, we add another degree of freedom to the system
by increasing $c$ and continue with the optimization from the last local
minimum until it converges again. We keep on increasing the number of
communities until we find that the newly introduced community does not improve
the overall community structure of the network (after the algorithm has settled
down again in a minimum). The community structure is assessed by the
fuzzification of the modularity function. The modularity, originally introduced
in \cite{newman04}, defines how good a community structure is by evaluating the
difference between the observed intra-community edge density and the expected
one based on a random graph model conditioned on the degree sequence of the
network. In a random graph with exactly the same degree sequence as the original
graph, the probability of the existence of an edge between vertices $i$ and $j$
is $k_i k_j / 2m$, where $k_i$ is the degree of vertex $i$ and $m$ is the total
number of edges in the network. The original, ``crisp'' modularity of a network
with vertex $i$ belonging to community $c(i)$ is then defined as:

\begin{equation}
Q = \frac{1}{2m} \sum_{i,j} \left[ A_{ij} - \frac{k_i k_j}{2m} \right] \delta_{c(i), c(j)}
\label{eq:modularity}
\end{equation}

where $\delta_{c(i), c(j)}$ is 1 if vertex $i$ and $j$ belong to the same
community ($c(i) = c(j)$), 0 otherwise. Since the community structure
in our algorithm is not clear-cut, the following predicate: ``vertex $i$ and
$j$ belongs to the same community'' also has a fuzzy truth value between
0 and 1. When the membership degree $u_{ki}$ is considered the probability
of the event that vertex $i$ is in community $k$, the probability of the event
that vertex $i$ belongs to the same community as vertex $j$ becomes the dot
product of their membership vectors, resulting in the already introduced
similarity measure $s_{ij}$, which can be used in place of $\delta_{c(i), c(j)}$
to obtain a fuzzified variant of the modularity:

\begin{equation}
Q_f = \frac{1}{2m} \sum_{i,j} \left[ A_{ij} - \frac{k_i k_j}{2m} \right] s_{ij}
\end{equation}

Note that in the case of crisp communities (there exists only one $k$ for every
vertex $i$ such that $u_{ik} = 1$), the fuzzified modularity $Q_f$ is exactly
the same as the crisp modularity $Q$. In order to determine the optimal number
of fuzzy communities, we iteratively increase $c$ and choose the one which
results in the highest fuzzified modularity $Q_f$.

\subsection{Parametrization of similarity and dissimilarity constraints}

Next, we discuss the appropriate choice of the remaining parameters ($\mx{W}$
and $\mx{\tilde{S}}$). These parameters are not
crucial to the final result of the algorithm, but they provide a way to inject
additional \emph{a priori} knowledge into the algorithm. Note that the goal
function (see Eq.~\ref{eq:goal_function}) is a weighted sum of the
difference between the desired and the calculated similarity values. The
algorithm tries to minimize the difference by fitting the membership values in
an appropriate way. Without any further \emph{a priori} knowledge,
$\mx{\tilde{S}}$ is the adjacency matrix $\mx{A}_G$
and $\mx{W}$ is a matrix containing only ones. This means that the dot product
of the membership vectors define the similarities, and the desired similarity
is 1 for adjacent and 0 for nonadjacent vertices, stating that the endpoints of
the edges should be as similar as possible, while keeping disconnected edges
dissimilar. The latter requirement is important: if we would only specify that
the endpoints should be similar for connected vertex pairs, the algorithm would
converge to a state where every vertex belongs to the same community.

Depending on the domain from which the network being analyzed originates, there
may be some additional knowledge about the original mechanism that created the
network, or there may be some uncertainty in the data. $\mx{W}$ can be used to
fine-tune the algorithm by making use of the domain-specific knowledge. The
general purpose of $w_{ij}$ is to emphasize the connections where the calculated
similarity should match the expected one and skip the connections where it is
hard or impossible to specify an expected similarity. $w_{ij}$ can also be useful
in the analysis of weighted networks.

\begin{figure}[tb]
\includegraphics[width=0.3\textwidth]{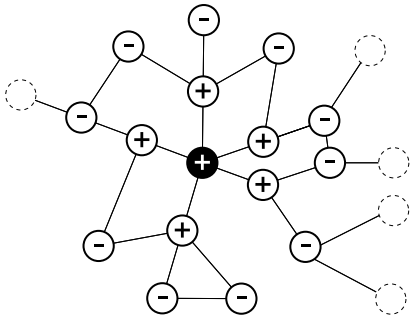}
\caption{The idea of distance-based relaxation. Direct neighbors of vertex $i$
(denoted by plus signs) are assumed to be similar to vertex $i$ (shown in black):
$\tilde{s}_{ij} = 1, w_{ij} > 0$. Vertices at most $k$ steps far from vertex
$i$ that are not direct neighbors (denoted by minus signs) are assumed to be dissimilar:
$\tilde{s}_{ij} = 0, w_{ij} > 0$. Vertices being farther than $k$ steps
(denoted by dashed circles) have no similarity specification with respect to vertex $i$:
$w_{ij} = 0$. The figure illustrates the case of $k=2$.
\label{fig:distance_relaxation}}
\end{figure}

Consider a large friendship network as an example. In a friendship network, a
reasonable assumption is that an existence of a connection between $A$ and $B$
predicts some kind of similarity between them. However, a missing connection
between $A$ and $B$ does not necessarily mean dissimilarity, it might happen
that $A$ and $B$ did not have a chance to meet and form a connection! To account
for this, one can assume that $A$ is similar to its direct neighbors and dissimilar
to its second order neighbors only (because they were likely to meet through
their common acquaintances). If necessary, this assumption can be incorporated into
Eq.~\ref{eq:goal_function} by setting the weight of the connections of
$A$ beyond its second order neighbors to zero. We call this modification the
\emph{distance-based relaxation} of the model. For an illustration of the concept,
see Fig.~\ref{fig:distance_relaxation}.

The proper choice of $w_{ij}$ also allows us to analyze the community structure
of networks with incomplete data. An example of this kind of a network is
described in \cite{negyessy06}. A graph model of the visuo-tactile cortex of
the macaque monkey was built based on the neural connections of the brain areas
already documented in the literature. However, there are actually two kinds of
missing edges in this network: the absence of an edge between brain areas $A$
and $B$ can either mean that the specific connection was tested for
experimentally and found to be nonexistent or that the connection has not ever
been sought for at all (due to, e.g., methodological difficulties). Our model can
account for this difference by setting the weight of the suspected connections
to zero and checking the similarity of the vertices involved after the analysis.
We will discuss this later in Section \ref{results}.

We also note that several other similarity measures can also be used when
one defines the expected similarity matrix $\mx{\tilde{S}}$ as long as
the similarities are normalized into the range $[0, 1]$. Based only on the
neighborhood of vertices, one can use the \emph{cosine similarity}
\cite{salton83} or the \emph{Jaccard similarity index} \cite{jaccard1901}.
More sophisticated, matrix-based methods have been studied in the papers of
Jeh and Widom \cite{jeh02}, Blondel \emph{et al} \cite{blondel04} and
Leicht \emph{et al} \cite{leicht06}. Some of these measures are not normalized
to the range $[0, 1]$, but this can be done easily by using an appropriate
transformation (e.g., dividing the similarities by the largest one found in
the network).

\section{Benchmarks and applications}

\label{results}
Generally, the community structure of a network is not uniquely defined.
Several partitions might exist that approximate the underlying structure
equally well, especially if the network exhibits an overlapping or hierarchical
community structure. As shown in \cite{palla05}, overlapping communities are
present in many networks ranging from co-authorship networks to protein
interactions. We expect our algorithm not only to discover this overlapping
structure but also to exactly quantify the membership degree of each vertex in
all of its communities.

Unless stated otherwise, we parametrized our algorithm as follows: $w_{ij} = 1$
for all $i, j$ and the desired similarity $\tilde{s}_{ij}$ was 1 if vertices
$i$ and $j$ were connected or $i$ was equal to $j$, 0 otherwise. The
automatic selection of the bridges was achieved by the standardization of the
bridgeness scores: a vertex was considered a bridge if its bridgeness score
was at least one standard deviation higher than the mean bridgeness score of
the vertices of the network.

\subsection{Benchmarks on computer-generated graphs}

\begin{figure}[tb]
\includegraphics[width=0.45\textwidth]{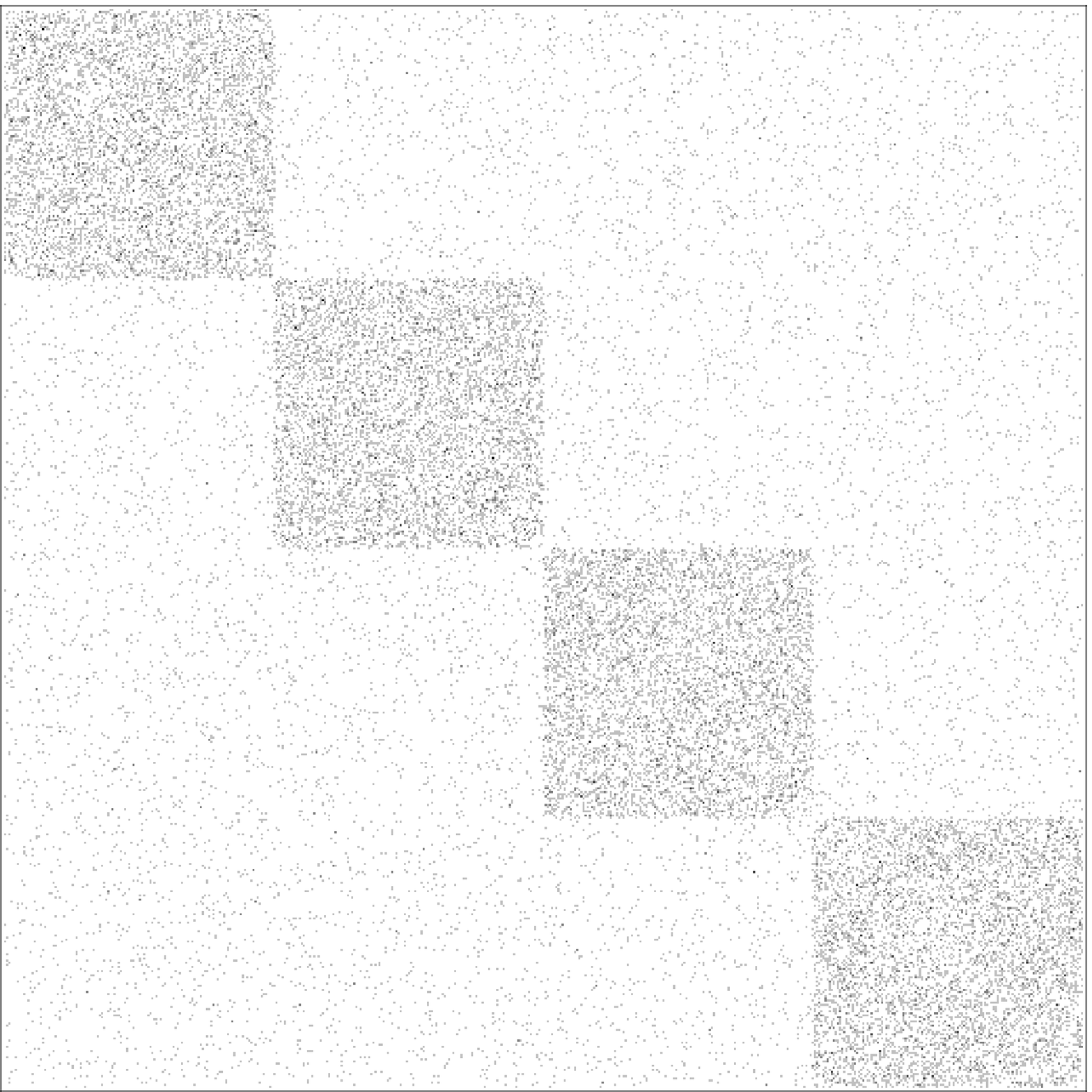} \\
~\\
\includegraphics[width=0.45\textwidth]{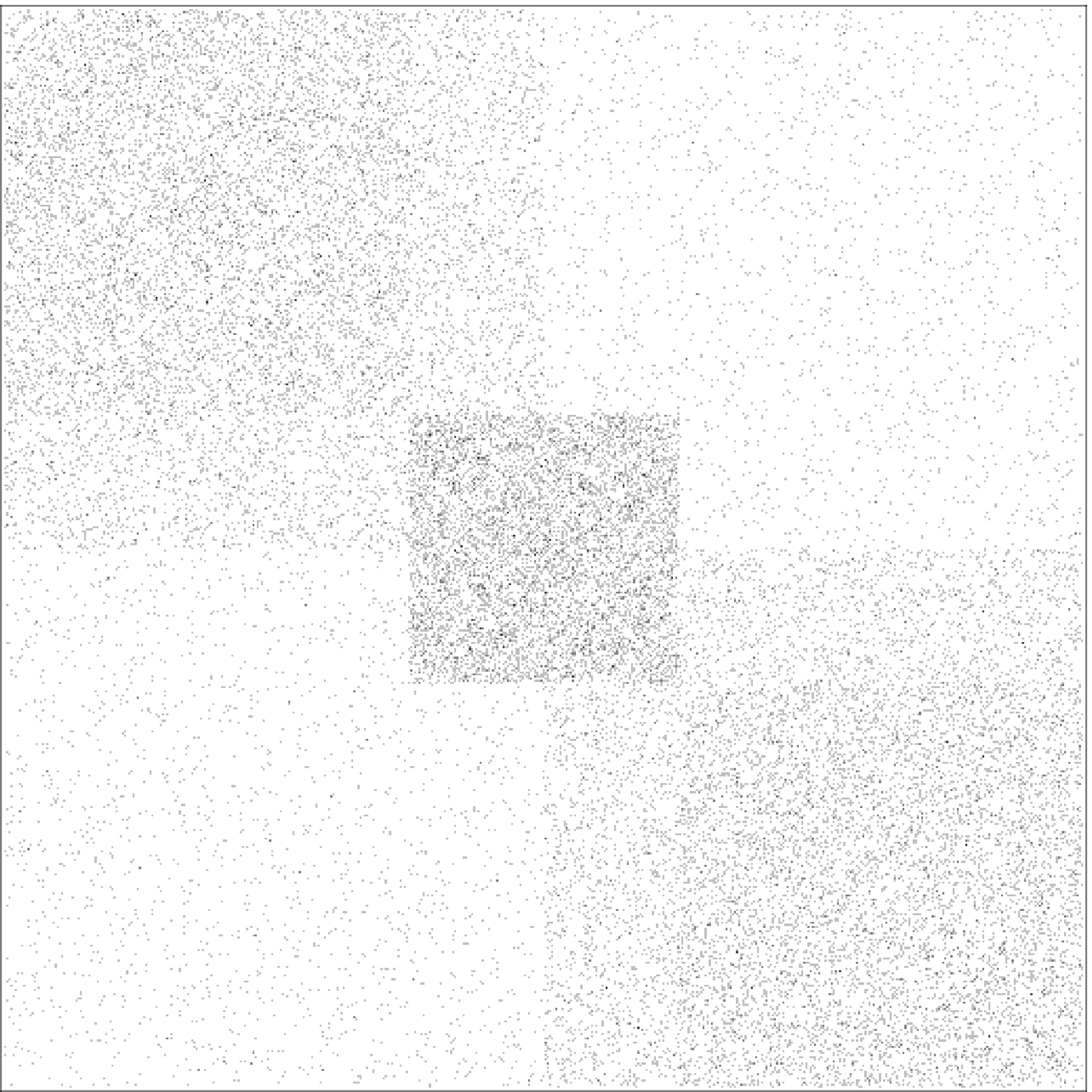}
\caption{\label{fig:adjacency}
Adjacency matrices of graphs with nonoverlapping (top) and overlapping
(bottom) community structure we used for benchmarking our algorithm.}
\end{figure}

We tested our method on several computer-generated networks with nonoverlapping
and overlapping community structure as well. Nonoverlapping community
structures were generated on graphs with 1024 vertices grouped into four
communities, each containing 256 vertices. Each vertex had an average of
$k_{in}=24$ links to other vertices in the same community and an additional
$k_{out}=8$ links to vertices from different communities. The generated graph
had 16,384 edges and a density of 0.031. Overlapping communities were
introduced by grouping the vertices into two communities and declaring 128
vertices in both communities as bridge vertices. Regular vertices kept their
connectional patterns, having 24 links on average to other vertices in their
community and 8 links to the other community. Bridge vertices had 6 links to
other vertices in their community, 12 links to other bridge vertices in their
community, 6 links to bridge vertices of the other community and 8 links to
regular vertices of the other community. The edge count and the density was
equal to the nonoverlapping case. Fig.~\ref{fig:adjacency} shows a possible
adjacency matrix for both the nonoverlapping and the overlapping case.

\begin{figure}[tb]
\subfigure[~]{
\label{fig:bmark}
\includegraphics[width=0.22\textwidth]{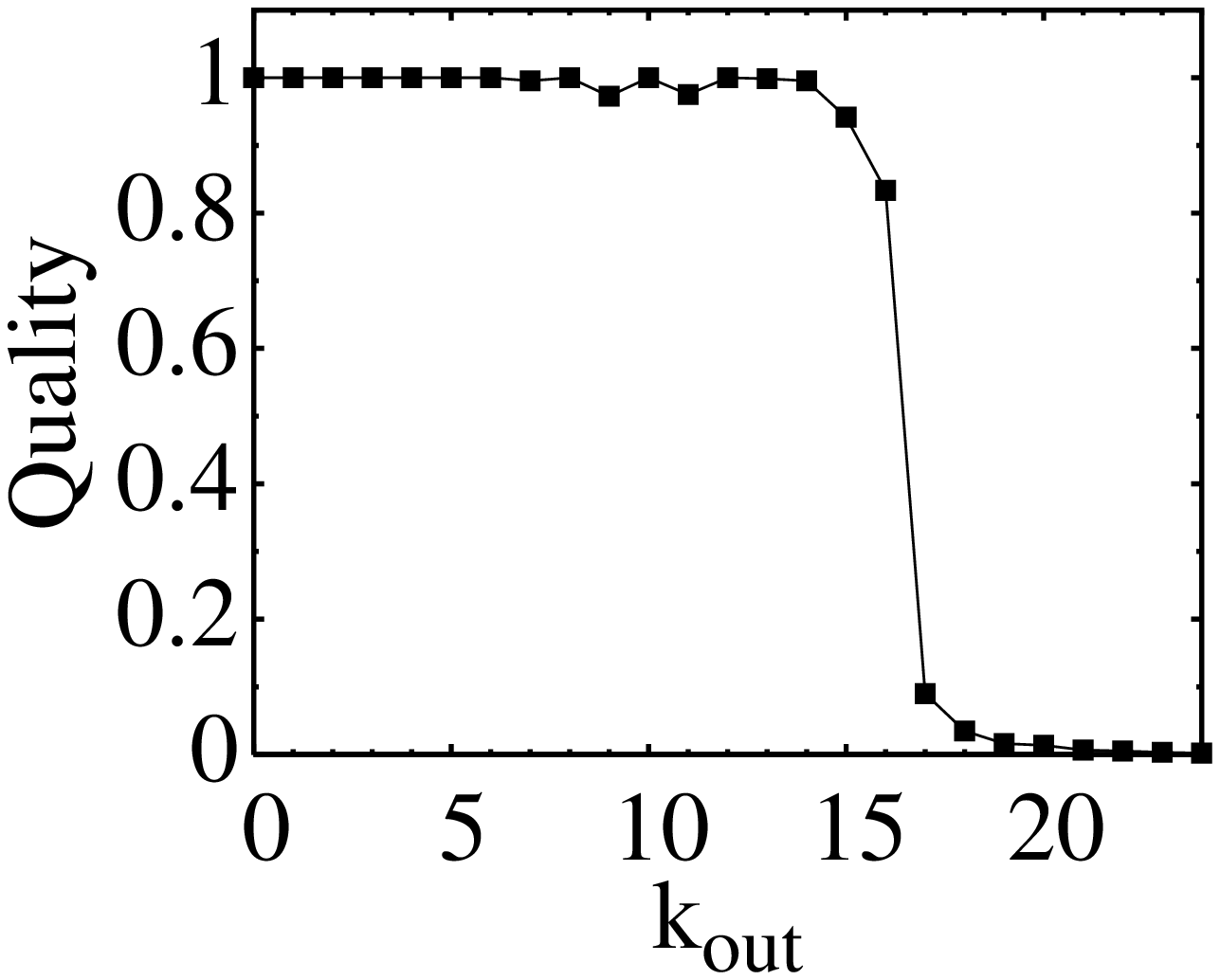}
}\hfill
\subfigure[~]{
\label{fig:bridgenesses}
\includegraphics[width=0.22\textwidth]{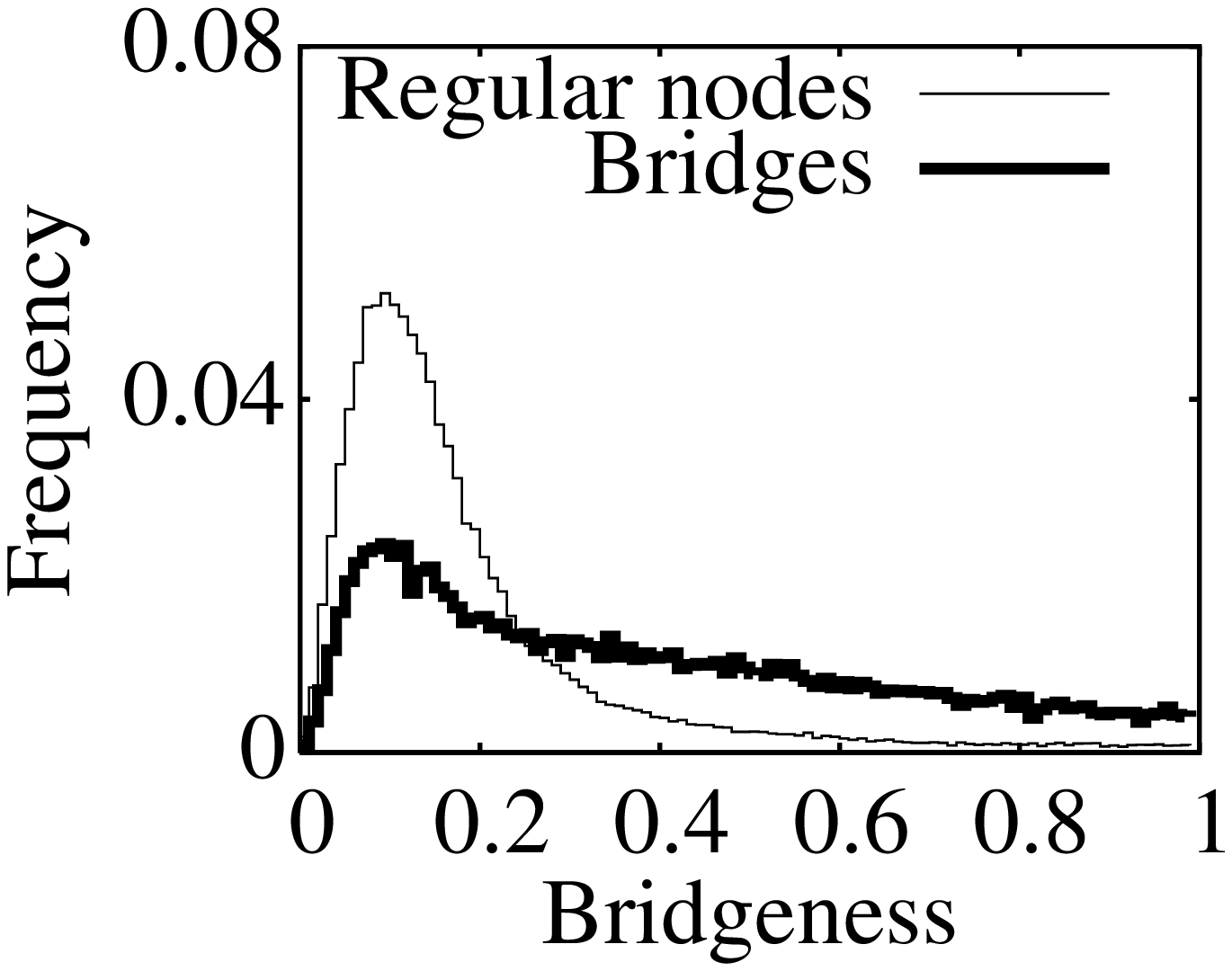}
}
\caption{Panel $(a)$ shows the performance of the algorithm for a graph with
nonoverlapping community structure. Inter-cluster link count ($k_{out}$) was
varied while keeping the average degree ($k_{in}+k_{out}$) constant.
The quality of the obtained result was measured using the normalized mutual
information of the found and real communities \protect\cite{danon05}.
Panel $(b)$ shows the frequencies of bridgeness values in a graph with
overlapping community structure. Thin line shows frequencies for regular
nodes, thick line shows frequencies for bridge nodes. The bin width of the
histogram was set to 0.01 (100 bins). }
\end{figure}

In order to compare a fuzzy partition with an expected hard partition, we
introduced the notion of \emph{dominant community}. The dominant community of a
vertex is the community to which it belongs to the greatest extent. Formally,
community $i$ is the dominant community of vertex $j$ if $u_{ij} \ge \max_k
u_{kj}$ for $1 \le k \le c$. Out of 1000 graphs with nonoverlapping community
structures, the algorithm classified all vertices correctly in 97.4\% of the
test cases after converting the achieved fuzzy partition to its hard
counterpart using the dominant communities. It was also able to infer the
actual number of communities automatically in all cases using the fuzzified
modularity. To further study the distribution of intra-community and
inter-community edges, we varied the number of inter-community edges ($k_{out}$)
from 0 to 24 while keeping $k_{in}+k_{out}$ constant. When $k_{out}$ reaches 24,
the graph practically becomes an Erd\H{o}s-R\'enyi random graph devoid of any
community structure, since the connectional probability between any two of the
pre-defined communities is equal.  Fig.~\ref{fig:bmark} shows the results of
the benchmark. The quality of the calculated community structure was assessed
by the normalized mutual information as described in \cite{danon05}.
Interestingly, the performance of
the algorithm degrades suddenly when the number of inter-community links
exceeds 16. This is the point where on average there are more links between the
communities than inside them.

Generated graphs with overlapping community structure were used to test the
sensitivity of the algorithm to vertices standing between communities. The
model we used declared 128 vertices out of 512 in both communities as
bridge candidates, and clearly distinguished them by their different
connectional patterns: bridge candidates tended to connect to each other with a
higher probability than to the regular vertices in their communities, even if
they originally belonged to different communities, creating an
overlap between the two communities. Because of the randomized nature of this
model, not all bridge candidates became real bridges between the communities,
but they had a significantly higher chance of becoming one. We used the bridgeness
value introduced in Section \ref{bridgeness} to assess the quality of the results.
We expected that bridge candidate vertices exhibit a different bridgeness score
distribution than the regular vertices in the same graph. We also required that
vertices identified as bridges by our algorithm should be among those that have
been declared bridge candidates before test graph generation. We generated 1000
random graphs using this graph model and plotted the distribution of the
bridgeness scores on Fig.~\ref{fig:bridgenesses}.  The different nature of the
two distributions was supported by a Kolmogorov-Smirnov test (p-value less than
$2.2 \times 10^{-16}$). Regular vertices usually had lower bridgeness scores
than the bridge candidates, and we found that 92.8\% of the identified bridges
(based on their standardized bridgeness scores) were among bridge
candidates, confirming that the algorithm is sensitive to the existence of
overlaps between communities.

\subsection{Social and collaboration networks}

\begin{figure}[tb]
\subfigure[~]{\label{fig:uk1}\includegraphics[width=0.22\textwidth]{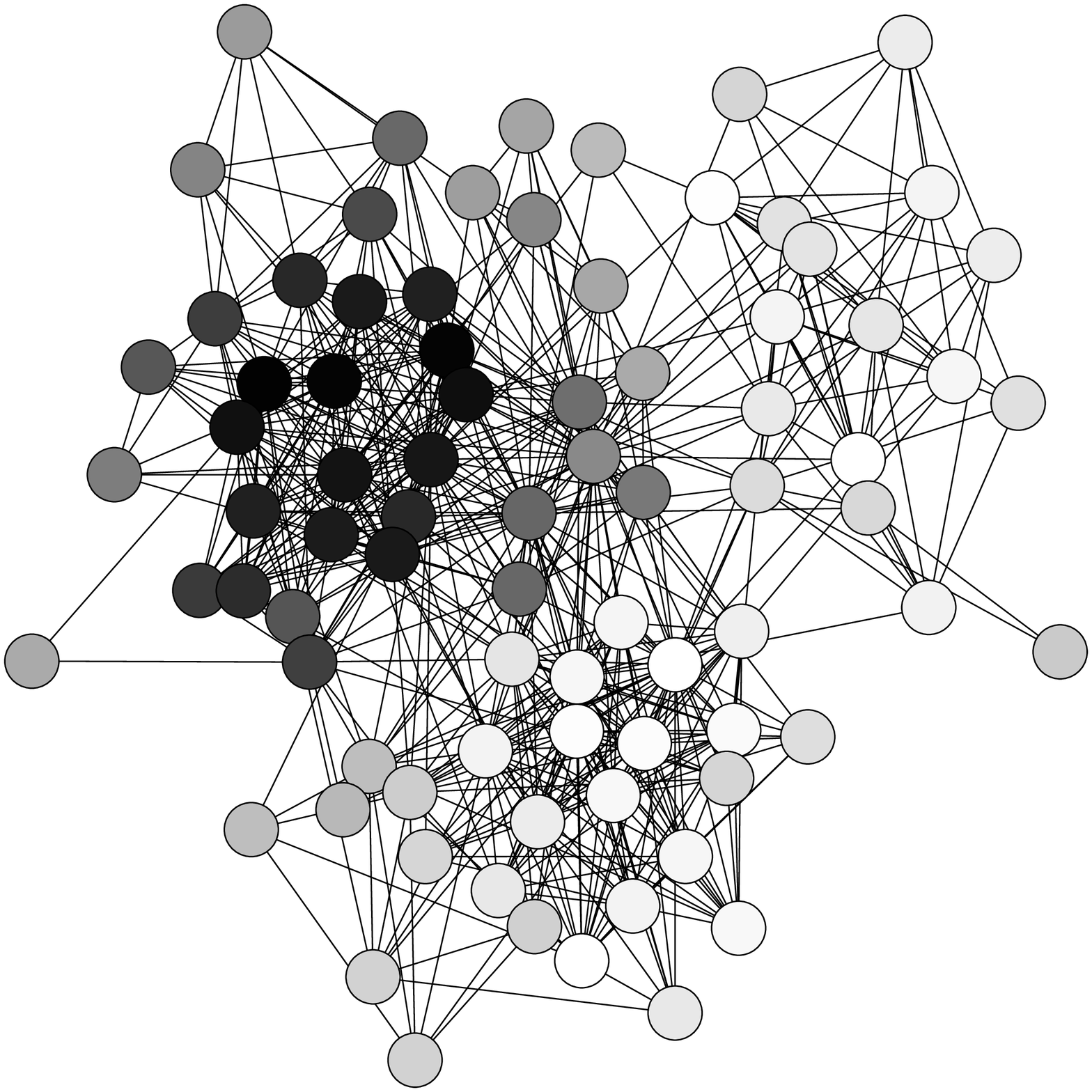}} \hfill
\subfigure[~]{\label{fig:uk2}\includegraphics[width=0.22\textwidth]{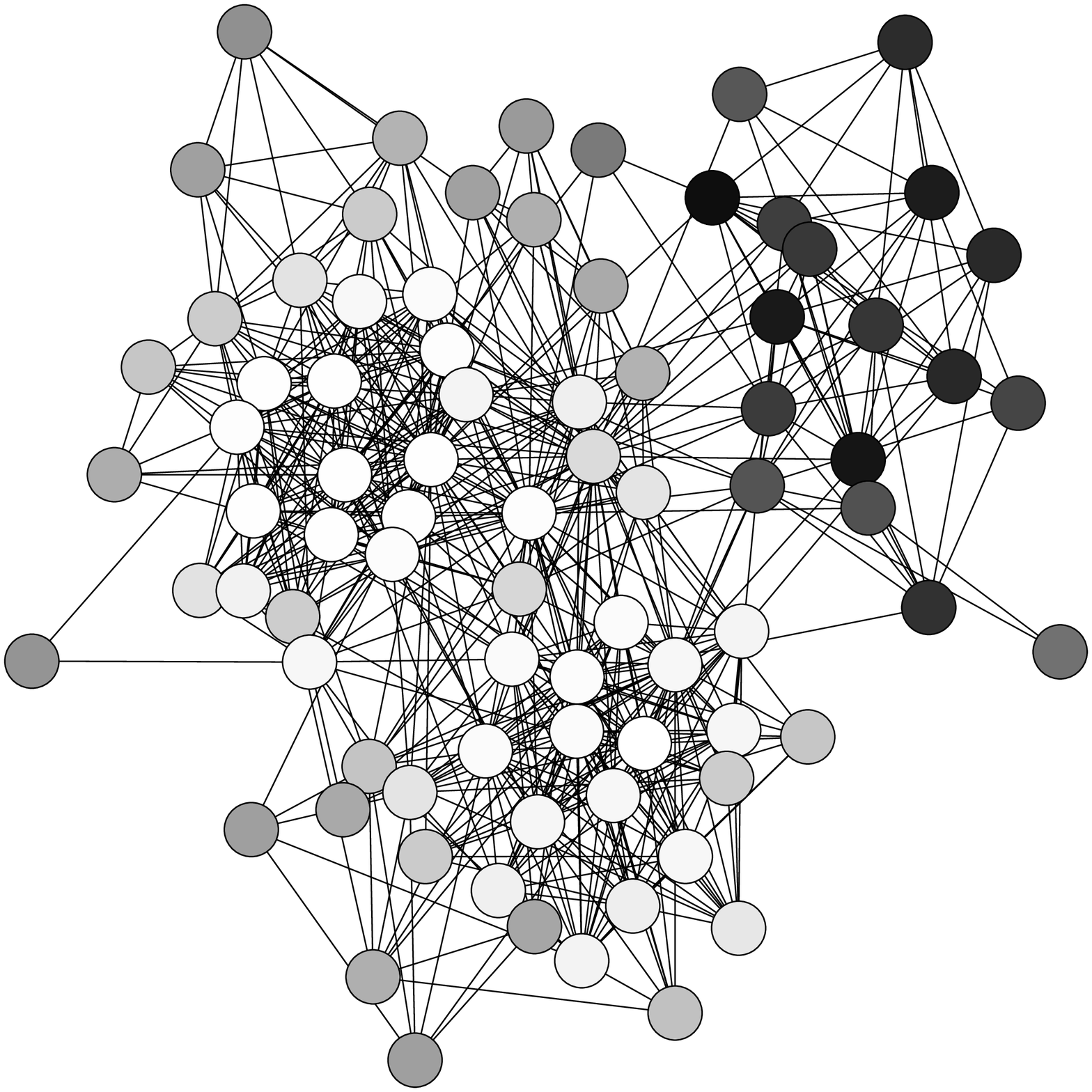}} \\
\subfigure[~]{\label{fig:uk3}\includegraphics[width=0.22\textwidth]{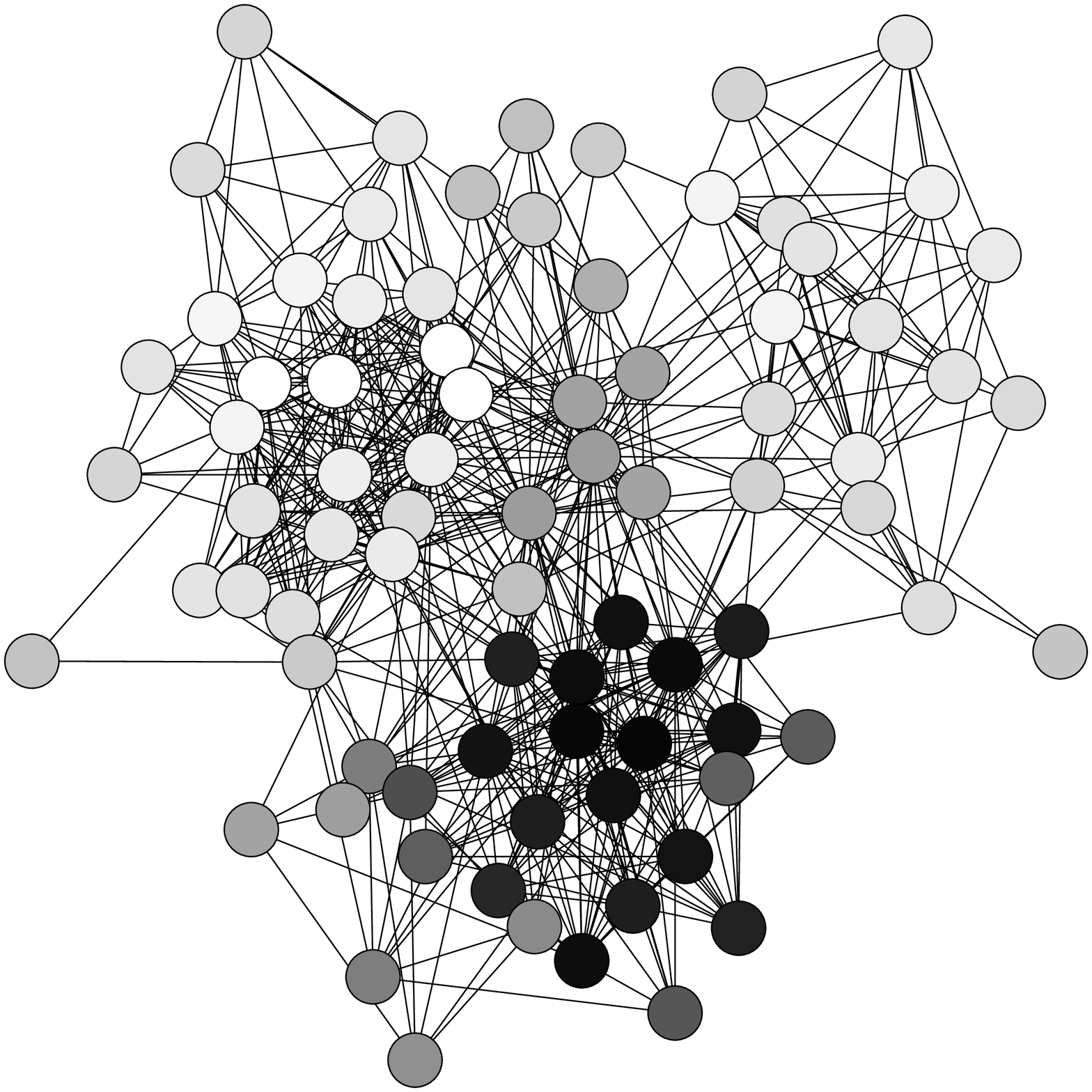}} \hfill
\subfigure[~]{\label{fig:uk4}\includegraphics[width=0.22\textwidth]{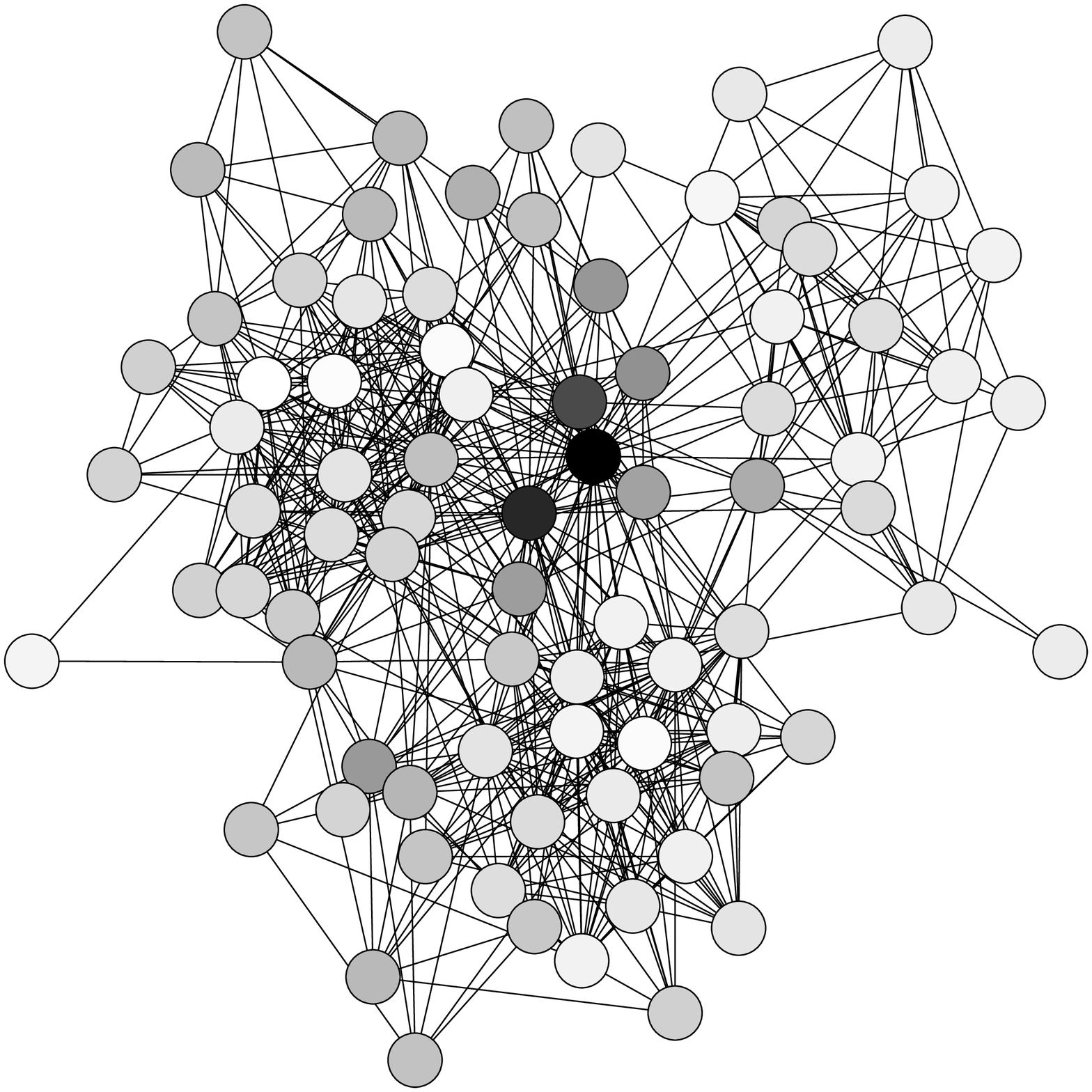}}
\caption{The fuzzy communities of the UK university faculty dataset. Panel (a), (b)
and (c): vertices colored according to the membership functions of community 1, 2 and
3, respectively. Darker shades represent larger membership values. Panel (d):
vertices colored according to the degree-corrected bridgeness scores. Darker shades
represent higher bridgeness.\label{fig:uk}}
\end{figure}

To evaluate the performance of our method on a real dataset, we used the social
network of the academic staff of a given Faculty of a UK university consisting
of three separate schools. The network structure was constructed from
tie-strength measured with a questionnaire, where the items formed a reliable
scale. Reliability was assessed by Cronbach's $\alpha$ \cite{cronbach51}. Our
questionnaire achieved a Cronbach's $\alpha$ of 0.91, suggesting high internal
consistency and reliability. The questionnaire was completed by every
member of the academic staff. In this study, we used the personal friendship
network, ignoring the directionality and the weight of the edges. A fuzzy
community detection for three communities was performed on the graph. To show
the results in grayscale, we decided to draw three individual figures
(Fig.~\ref{fig:uk1}, \ref{fig:uk2} and \ref{fig:uk3}), showing the values of
the membership functions for community 1, 2 and 3, respectively, using
different shades of gray as fill colors for the vertices. Fig.~\ref{fig:uk4}
shows the degree-corrected bridgeness values for each vertex. Other centrality
measures resulted in the same corrected bridgeness scores after normalization.

This dataset also contained explicit information regarding the expected
community structure, since for every node, we knew which school in the Faculty
does it belong to. We defuzzified the results using the dominant communities for
every vertex. The defuzzification revealed that all crisp communities consisted of
almost exclusively the members of a single school inside the Faculty. 75 out of
81 vertices were classified correctly, 4 were misclassified (and all of them had
a bridgeness value greater than 0.7), and there were 2 vertices for which no
expectation was given because of lack of information in the questionnaire. It
is also noteworthy that the maximal fuzzy modularity ($Q_f$ = 0.2826) was reached at
$c=6$, suggesting further subdivisions of the schools, although the improvement
of the modularity compared to the case of $c=3$ ($Q_f$ = 0.2541) was not significant.

\begin{figure}[tb]
\subfigure[~]{\includegraphics[width=0.22\textwidth]{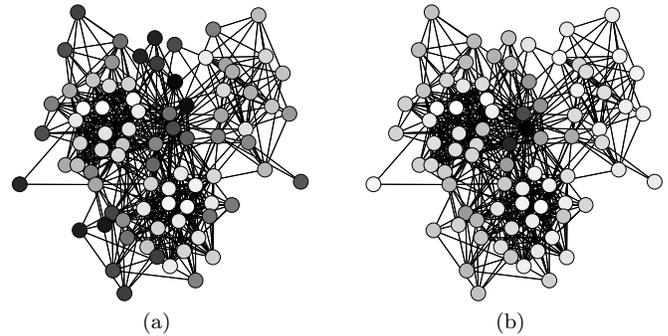}}
\hfill
\subfigure[~]{\includegraphics[width=0.22\textwidth]{uk_br.eps}}
\caption{\label{fig:bridgeness_comparison}Comparison of the uncorrected (left)
and degree-corrected bridgeness scores (right) in the UK university dataset. Vertices are
colored according to their respective bridgeness scores. Darker shades represent higher
bridgeness scores. Note how the uncorrected bridgeness score correlates with the
centrality of the vertices in their respective community.}
\end{figure}

Degree-corrected bridgeness scores for $c=3$ (Fig.~\ref{fig:uk4}) are particularly
interesting. Highly scored individuals belong to all three communities
at the same time to some extent, maintaining connections to all of them. On the
other hand, vertices with low degree-corrected bridgeness scores can be
thought as the cores of the communities. We also notice that the peripheries of
the communities also belong almost equally to all of the communities (note the
similar grey shades in Fig.~\ref{fig:uk1}, \ref{fig:uk2} and \ref{fig:uk3} for
these vertices), but the degree-corrected bridgeness scores suppress this
effect because of their low degree. The uncorrected and the degree-corrected
scores are compared side-by-side on Fig.~\ref{fig:bridgeness_comparison}. We
also point out that the uncorrected bridgeness scores can be used as a measure
of the centrality of a given vertex with respect to its own dominant community
by substracting it from 1.

The next dataset we studied was the co-authorship network of scientists working
on network theory and experiment, as published in \cite{newman06}. The network
consists of 1589 scientist and 2742 weighted, undirected connections. Edge
weights are derived from the number of joint publications: if author $A$ and
$B$ share a paper where they are both authors and the paper has $n$ total
authors, this contributes by $\frac{1}{n}$ to the total weight of the edge. We
extracted the giant component of the network consisting of 379 scientists and
914 connections and let our algorithm determine the number of communities using
the fuzzified modularity again. The optimum ($Q_f$ = 0.7082) was found with $c=30$
communities. The value of $c$ was confirmed by the visual inspection of
the eigenvalues of the Laplacian matrix. Without names, we observe
that vertices with the highest centralities according to our measure were
similar to the ones chosen by the community centrality measure introduced in
\cite{newman06} and mostly represented senior researchers of the field of
network science. Bridges were detected by standardizing the bridgeness values
and considering vertices with a z-score higher than 1 as bridges. The 31 bridge
vertices were mostly post-doctorate researchers who collaborated with more
than one senior researcher of the field.

\subsection{Cortical networks and the case of incomplete data}

\label{incomplete_data}
To test how our method performs on graphs with missing data (vertex pairs for which
no information regarding their connectedness was known), we used the graph model
of the macaque monkey's visuo-tactile cortex as published in \cite{negyessy06}.
The graph consists of 45 vertices representing brain areas, and 463 directed connections
representing neuronal pathways between the areas. Disconnected vertices do not
necessarily mean that there is no connection between them: some of them have been
explicitly tested for and found to be absent, others have simply not been tested for
(but generally thought to be absent), and there are 13 vertex pairs in total where
neuroanatomists strongly suspect that there exists a connection between them. The
graph itself consists of two distinct and mostly nonoverlapping communities
corresponding to the visual and the somatosensory cortex. Other, anatomically
meaningful subdivisions of the cortices (like the dorsal and the ventral stream in
the visual cortex) are known as well. We also note that 11 out of the 13 suspected
connections are heteromodal in the sense that they go between the visual and the
somatosensory cortex.

\begin{figure}[tb]
\includegraphics[width=0.45\textwidth]{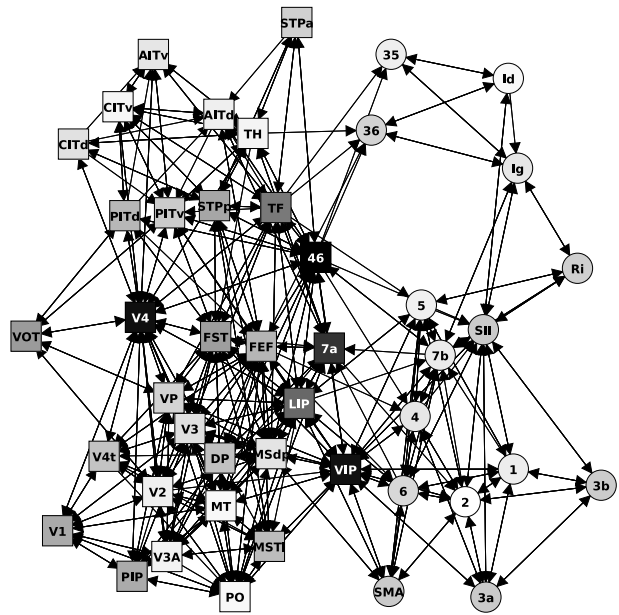}
\caption{\label{fig:db_cortex} The cortical network dataset
\protect\cite{negyessy06}. Rectangular vertices are
visual areas, circular vertices are somatosensory areas. Vertices are colored
according to their degree-corrected bridgeness values for $c=4$. Detected
bridges are highlighted with white text color.
}
\end{figure}

\begin{figure}[tb]
\includegraphics[width=0.45\textwidth]{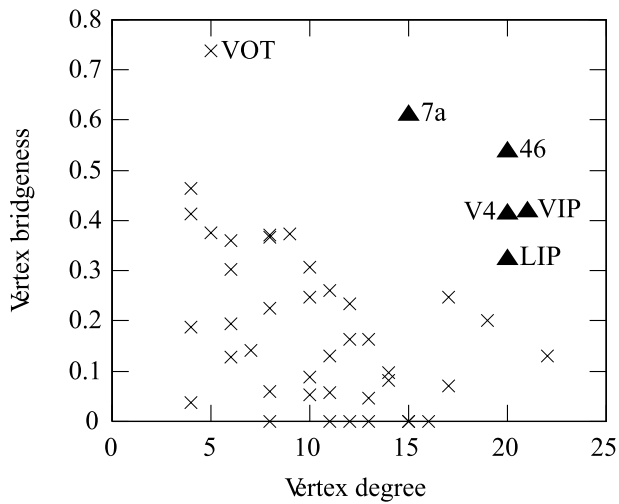}
\caption{\label{fig:cortex_degree_br_plot} The degree-bridgeness plot of the
vertices of the cortical network dataset for $c=4$. Crosses denote regular
vertices and triangles denote bridges. Bridges and the single significant
outlier (area $VOT$) have also been marked with the name of the
corresponding area. The remaining names were omitted for sake of clarity.
$VOT$ is a biologically relevant example of a vertex with low degree and
high bridgeness.}
\end{figure}

To account for the uncertainty and the directedness of the edges in the graph,
we specified $w_{ij}$ as follows: $w_{ij}$ was 0 if there was a nonreciprocal
connection between area $i$ and $j$ (area $i$ connected to $j$, but no pathway
was found in the reverse direction) or if the connection was one of the
suspected ones, otherwise $w_{ij}$ was 1. The optimal fuzzy modularity (0.2766)
was reached at $c=4$. We examined the results for $c=2$ and $c=4$. The
case of $c=2$ classified the nodes correctly: all of the
somatosensory areas were associated with the somatosensory cortex and most of
the visual areas were associated with the visual cortex, except a few areas
with a surprisingly high bridgeness (over 0.85). The vertex with the highest
bridgeness (0.99) was area $46$, a part of the dorsolateral prefrontal cortex,
and it does not have functions related to low-level sensory information
processing. Area 46 is rather a higher level (supramodal) area, which plays a
role in sustaining attention and working memory, and being a bridge between the
visual and the somatosensory cortex, it integrates visual, tactile and other
information necessary for the above mentioned cognitive functions.  Other
relevant bridges found with $c=4$ were area $VIP$ (where the literature has already
suggested that it should be split into two areas $VIPm$ and $VIPp$, which
establish stronger connections with visual or sensorimotor areas, respectively
\cite{lewis00}), $LIP$, $V4$ and $7a$. $VIP$ and $LIP$ are involved with hand
and eye coordination, respectively, and both of these functions require combined
information from visual and tactile signals as well. Area $7a$ integrates
visual, tactile and proprioceptive signals. Area $V4$ was defined originally as
the human color center \cite{lueck89,mckeefry97}, while it was also suggested
that a separated ensemble of $V4$ neurons successfully encode complex shapes
based on the curvature of the shape boundaries \cite{pasupathy06}. The
functional heterogenity is in accordance to the subdivision of $V4$ into
different regions as suggested by Bartels \emph{et al} \cite{barthels00}.  We
conclude that the bridges we found are in concordance with the assumed higher
level roles of these areas. Fuzzy community detection for $c=4$ was also able
to separate the dorsal and the ventral stream of the visual cortex, only area
$7a$ and $VIP$ were misclassified, but they retained their bridgelike
properties as well as area $46$. The degree-corrected bridgeness values for
$c=4$ are shown on Fig.~\ref{fig:db_cortex}. Plotting the uncorrected bridgeness
values versus a chosen centrality measure (in our case, the vertex degree),
shown on Fig.~\ref{fig:cortex_degree_br_plot} was found to be a useful visual
aid for separating bridge vertices and outliers.

To approximate the probability of the suspected connections, we calculated the
pairwise similarities of the vertices involved and considered the similarity as
the probability of the existence of a connection. This is based on the idea
that one can consider the membership value $u_{ij}$ as the probability of
vertex $j$ belonging to community $i$. In this sense, the similarity of
vertices $i$ and $j$ is the probability of the event that they are in the same
community, and according to our prior assumption that similarity implies
connectivity, we can think about higher similarity values as precursors for
existing connections. Without going into further details and possible
neuroanatomical implications, we concluded that all supposed connections of
area $LIP$ are less likely than the supposed connections of $VIP$, and among
the possible unknown connections of $VIP$, the connections with areas $4$ and
$6$ are the most probable.

\subsection{Comparison with other overlapping community detection algorithms}

In order to compare our method with earlier attempts on tackling the problem
of overlapping communities, we examined the CPM algorithm of Palla \emph{et al}
\cite{palla05}, the spectral method of Capocci \emph{et al} \cite{capocci05}
and the fuzzy method of Zhang \emph{et al} \cite{zhang07}. We tested all three
methods on the example graph shown on Fig.~\ref{fig:toy_graph} and on the
macaque monkey dataset introduced in Section \ref{incomplete_data}. For
the CPM algorithm, we used the original implementation published by the authors
at \texttt{http://www.cfinder.org}. The algorithm of Zhang \emph{et al} had a
weight exponent $m$ controlling the degree of fuzzification, but since the authors
provided no clue about the suggested value of the parameter, we used $m=2$,
which is the most typical choice of this parameter in other known applications of
the fuzzy $c$-means algorithm \cite{bezdek81}.

The proper community structure of the example graph was detected by all algorithms
we considered (including ours), although the spectral method of Capocci
\emph{et al} had to be tested on a different example graph with three cliques
(each of size 4) and a single connector node, because in the case of only two
communities, the only eigenvector that carries useful information is the first
nontrivial one, rendering correlation calculations meaningless. Moreover, the
global community structure became evident only after proper rearrangement of the
community closeness matrix provided the algorithm. The bridge-like property of
the connector vertex was inferred from the zero community closeness
values to all other vertices. The method of Zhang \emph{et al} and our method
produced the proper expected partition matrix with all the vertices except
vertex 5 classified strictly to one community or the other, while vertex 5
belonging to both at the same time with a membership degree of 0.5. The method
of Palla \emph{et al} identified vertex 5 as an outlier vertex, but after
adding more edges to it, it became an overlap between the communities.

The community structure of the cortical graph seemed to be a harder problem
for the algorithms. The method of Palla \emph{et al} failed to discover the
subdivision of the two main communities, only the visual and the somatosensory
cortex was discovered when we used a clique size of 5. Larger clique sizes
resulted in the discovery of the cores of the two communities, but we were
not able to recognize the subdivision of the dorsal and the ventral stream
in the visual cortex. However, the algorithm identified three overlaps
($V4$, $PITv$ and $TF$) for a clique size of 5 and two other overlaps
($LIP$ and $VIP$) for a clique size of 6. Three out of these five overlaps
were identified by our algorithm as well. The community closeness matrix
calculated by the method of Capocci \emph{et al} was harder to interpret,
but vertices $V4$ and $46$ clearly turned out to be bridges with zero
community closenesses to many other vertices. The method of Zhang \emph{et al}
was highly sensitive on the exact value of parameter $m$, classifying
40\% of the vertices as bridges for $m=2$. (Since the method provides a
membership matrix similar to ours, we used the standardized bridgeness
measure with a z-score threshold of 1). Lowering the weight
exponent to $m=1.3$ identified vertices $LIP$, $7a$ and $Ri$ as bridges.

We found that the results of our algorithm with respect to community
structure discovery and bridge identification do not contradict the results
of existing methods, and all the bridges found by our algorithm were classified
as bridges by at least one different method. The method of Capocci \emph{et al}
complements our algorithm especially well, since it discovers
local communities around a given vertex using the community closeness degrees
while our method provides useful insights into the global structure of the
network being analyzed, also indicating the presence of bridge vertices.

\section{Conclusion}

In this paper, we presented a fuzzy extension of classical community
detection algorithms based on the assumption that communities of complex
networks are formed by vertices with graded commitments towards at least one
community. Accordingly, every vertex is allowed to belong to multiple
communities with different membership degrees, represented by a single real
value $u_{ki} \in \left[0, 1\right]$ for each vertex $i$ and community $k$.
The $\mx{U} = \left[ u_{ki} \right]$ matrix encodes the membership values in
a compact form and allows us to define the similarities of the vertices as
$\mx{S} = \mx{U}^T \mx{U}$ in its simplest form. The similarities are then
optimized using gradient-based constrained optimization methods in order to
make connected vertices similar and disconnected vertices dissimilar. Based on
the results of the fuzzy community detection, we introduced a novel
concept called bridgeness, which can be used to measure to what extent is a
given vertex shared between the communities. Vertices with high bridgeness
values were shown to be important in various complex networks, including (but
not limited to) social networks, scientific collaboration networks and cortical
networks. A transformed variant of bridgeness can be used as a centrality
measure with respect to the dominant communities of a vertex.

We emphasize that this algorithm is expected to be highly useful in the
analysis of relatively small datasets (up to the magnitude of a thousand
vertices). The reason is that the algorithm assumes that every vertex has the
possibility to connect to all other vertices, and if they do not connect, they
do that because they are of no use to each other. In very large networks, this
assumption is not always realistic. However, the distance-based relaxation
introduced in Section \ref{distance_based_relaxation} can still be used in
these cases to account for the upper bound imposed on the distance of the
potentially interacting vertices.

\begin{acknowledgments}
The authors are grateful to Curt Koenders, Zolt\'an Somogyv\'ari and the
two anonymous reviewers for their useful comments on earlier versions
of this paper.
The UK academic staff social network study was made possible with the support
from the British Academy Small Grant Scheme, Grant Number: SG-42203. L\'aszl\'o
N\'egyessy was supported by the J\'anos Bolyai Research Scholarship of the
Hungarian Academy of Sciences.
\end{acknowledgments}

\bibliography{fuzzyclust}

\end{document}